\documentclass[aps,prl,twocolumn]{revtex4}
\usepackage{graphicx}
\usepackage{amsmath}
\usepackage{amssymb}
\usepackage{braket}
\usepackage{hyperref}
\usepackage{subfigure}
\usepackage[usenames,dvipsnames]{color}

\usepackage{graphicx}
\usepackage{subfigure}

\newcommand{\ct}{\cite}
\newcommand{\la}{\lambda}
\newcommand{\bi}{\bibitem}
\newcommand{\be}{\begin{equation}}
\newcommand{\ee}{\end{equation}}
\newcommand{\ba}{\begin{eqnarray}}
\newcommand{\ea}{\end{eqnarray}}

\newcommand{\noi}{\noindent}

\begin{document}

\title{Emergent topology and dynamical quantum phase transitions in two-dimensional closed quantum systems}
\author{Utso Bhattacharya and Amit Dutta \\
Department of Physics, Indian Institute of Technology, 208016, Kanpur}

\begin{abstract}
We introduce the notion of a dynamical topological order parameter (DTOP) that characterises   dynamical quantum phase transitions (DQPTs)  occurring  in the subsequent temporal
evolution  of {\it two dimensional} closed quantum systems, following a quench (or ramping) of a parameter of the Hamiltonian, {which generalizes the notion of DTOP introduced in Budich and Heyl,
Phys. Rev. B {\bf 93}, 085416 (2016) for one-dimensional situations}.  This DTOP is obtained  from the ``gauge-invariant" Pancharatnam phase
extracted from the Loschmidt overlap, i.e., the  modulus of the overlap between the initially prepared state 
 and its time evolved counterpart reached following a temporal evolution generated by the time-independent final Hamiltonian. This generic proposal is illustrated considering DQPTs occurring
 in the subsequent temporal evolution following a sudden quench of the staggered mass  of the topological  Haldane model on a hexagonal lattice where it stays  fixed to zero or unity and makes a discontinuous
 jump between these two values at critical times at which DQPTs occur.

\end{abstract}
\maketitle

Recent experimental advances in  realisation of closed condensed matter systems via cold atoms \ct{bloch08,lewenstein12,jotzu14} in optical lattices,  especially studies  of the real time evolution of closed quantum systems in cold atomic gases \ct{greiner02} have resulted
in an upsurge in related theoretical works. The experimental investigations focus on intriguing dynamical phenomena like   prethermalization \ct{kinoshita06,gring12,trotzky12},  light-cone like propagation of quantum
correlations \ct{cheneau12},  light-induced non-equilibrium  superconductivity and topological systems \ct{fausti11,rechtsman13} and many-body localization in disordered interacting systems \ct{schreiber15}. In parallel,   there have been numerous theoretical  attempts  to explore e.g., the growth of entanglement entropy following a quench \ct{calabrese06}, thermalization \ct{rigol08}, light-induced topological 
matters \ct{kitagawa10,lindner11}, dynamics of topologically ordered systems \ct{bermudez09,patel13,thakurathi13}, periodically driven closed
quantum systems \ct{mukherjee09,das10,Russomanno_PRL12,bukov16} and many body localization \ct{pal10,nandkishore15}.
{(For review, we refer to \ct{dziarmaga10,polkovnikov11,dutta15,eisert15,alessio16}.)} 


The proposal  of a DQPT was  put forward  by Heyl $et~ al.$ \ct{heyl13}, in close connection to  a  thermal phase transition in an equilibrium classical system;
the latter
can be detected, as proposed by Fisher \ct{fisher65} (See also, \ct{lee52,saarloos84}),  by  analyzing the zeros of the canonical partition function in a complex temperature plane (or
in a complex magnetic field \ct{lee52}). In DQPTs, on the other hand,  non-analytic behavior occurs at critical times in the subsequent real time evolution (following
 the quench)
 generated by the time independent final Hamiltonian; these  non-equilibrium transitions can be analyzed by locating the zeros of  the ``dynamical"  partition function generalized to the complex time plane.
 It is noteworthy that Fisher (or Yang-Lee) zeros have been experimentally observed within a central field where a qubit is coupled to the bath in such a way that it  experiences an effective
complex magnetic field \ct{peng16}.

Let us first elaborate on the basic notion of a DQPT focussing on the sudden quenching  of a one dimensional model and treating the overlap amplitude or the Loschmidt overlap (LO) as the analogue of the ``partition" function.
 Denoting  the ground state of the initial Hamiltonian as  $|\psi_0\rangle$ and the final Hamiltonian reached through the
quenching process as  $H_f$, the Loschmidt overlap is defined as  $G(t)=\langle\psi_0|e^{-iH_ft}|\psi_0\rangle$. Generalizing $G(t)$  to 
 $G(z)$ defined in the  complex  time ($z$) plane, one can introduce the notion of a dynamical free energy density, $f(z)=-\lim_{L\to \infty} \ln{G(z)}/L$, where $L$ is the linear dimension of the system. 
One  then looks  for the zeros of the $G(z)$ (or non-analyticities in $f(z)$) to define a dynamical phase transition.  For a transverse Ising chain, it has been observed \ct{heyl13}  that  when the system is suddenly quenched across the quantum critical point (QCP), the lines of  Fisher zeros  cross the imaginary time axis  at  instants $t_n^{*}$; at these instants  {the rate function of the return probability} defined as $I(t) = - \ln |G(t)|^2/L$ shows sharp non-analyticities.     Several subsequent studies \ct{karrasch13,kriel14,andraschko14,canovi14,heyl14,vajna14,sharma15,heyl15,palami15,vajna15,schmitt15,budich15,sharma16,divakaran16,huang16,puskarov16,zhang16,heyl16}  established that similar DQPTs are observed for  sudden quenches across the QCP for both integrable and non-integrable models. DQPTs have also been  observed when the final state that evolves with the final Hamiltonian is prepared through a slow ramping of a  parameter of the Hamiltonian \ct{sharma16,pollmann10}.
We note in the passing that the rate function $I(t)$  is related to the Loschmidt echo which has been
studied in the context of  equilibrium quantum phase transition and associated dynamics  \ct{quan06,rossini07,cucchietti07,venuti10,
sharma12,nag12,mukherjee12,dora13,zanardi07_echo,gambassi11,dorner12}.

These studies of DQPTs  have been generalized to two-dimensional systems \ct{vajna15,schmitt15} (where Fisher zeros usually form an area in the complex $z$-plane) and non-analyticities are manifested
in $I'(t)$, i.e.,  the time-derivative of $I(t)$ (and {\it not} in $I(t)$  itself). However, in the context of  the topological Haldane model, it has also been established that an effective one-dimensional behaviour  emerges when the 
quasi-momentum dependent Haldane mass term vanishes 
\ct{bhattacharya09} and  FZs form lines as in  the one-dimensional situation with non-analyticities in $I(t)$ itself.

Remarkably, {in Ref. [\onlinecite{budich15}], it was established  for the first time  that there exists a dynamical topological order parameter (DTOP) which can characterize DQPTs those occur following 
a sudden quench of one-dimensional integrable models; subsequently, in Ref. [\onlinecite{sharma16}], a similar DTOP  was proposed to characterize DQPTs those occur following slow quenches in similar models.}
This DTOP is expressed in terms of the Pancharatnam geometric phase (PGP) associated with the LO. The  existence of a DTOP  emphasizes the emergence of a topological structure associated with the temporal evolution of a quenched system. 
The essential and non-trivial question that we address in this paper is whether the DQPTs in two-dimensional models, reflected in the time derivative of the rate function $I(t)$, can also be characterized by
a DTOP obtained from the gauge-independent form of the Pancharatnam's phase extracted from the LO defined above. This question has gained importance following the experimental detection of DQPTs following a sudden quench of Haldane like models \ct{flaschner16}.

To address the question raised above, we refer  to a  two-dimensional  model,  which can be decoupled to two-level systems for each momentum mode ${\vec k}$ (for example, the topological Haldane model), and assume that the system is initially
prepared in the ground state ($|\psi_0 \rangle$) of the initial Hamiltonian $H_i(\la_i)$; the parameter $\la$ is suddenly changed from $\la_{i}$ to $\la_f$ while the state of the system stays frozen
in  the state $|\psi_0 \rangle$. For the wave vector $\vec k$, the state $|\psi_0 \rangle$ can  be decomposed in the eigenbasis of the final Hamiltonian $H(\la_f)$ as  $|\psi_{0_{\vec k}} \rangle = v_{\vec k} |1_{\vec k}^f\rangle  + u_{\vec k} |2_{\vec k}^f\rangle$, with $|u_{\vec k}|^2 + |v_{\vec k}|^2 =1$: here, $|1_k^f\rangle$ and $|2_k^f\rangle$ are the ground
 and the excited states of the Hamiltonian $H_{\vec k}(\la_f)$ with energy eigenvalues $-\epsilon_{\vec k}^f$ and $\epsilon_{\vec k}^f$, respectively. We shall then study the evolution of the state $|\psi_{0_{\vec k}}\rangle$  with the final Hamiltonian $H_{\vec k}(\la_f)$;
the Loschmidt overlap, $L_{\vec k} = \langle \psi_{0_{\vec k}}| \exp(- iH_{\vec k}(\la_f)t)|\psi_{0_{\vec k}} \rangle$ for the ${\vec k}$-th mode is immediately
written in the form 
\be
L_{\vec k} (t)= |v_{\vec k}|^2  \exp(i  \epsilon_{\vec k}^f t) + |u_{\vec k}|^2 \exp(-i \epsilon_{\vec k}^f t).
\label{eq_LO}
\ee
Generalizing to the complex time ($z$) plane and defining the dynamical partition function $G(z) = \prod_{\vec k} G_{\vec k} (z) = \prod_{\vec k} \langle \psi_{0_k}| \exp(- H_{\vec k}(\la_f)z)|\psi_{0_k} \rangle$ one can then immediately find the the Fisher zeros (i.e., the zeros  of $G(z)$)  given by:

\be
z_n({\vec k}) = \frac 1 {2 \epsilon_{k}^f}  \left[ \ln \left(\frac {|u_{\vec k}|^2 }{1-|u_{\vec k}|^2}\right) + i \pi(2n+1)\right],
\label{eq_fisher_zero}
\ee
where $n=0,\pm 1, \pm 2, \cdots $. These Fisher zeros form dense areas in the complex plane (in the thermodynamic limit) as long as they are  explicit functions  of $k_x$ and $k_y$. These areas (usually generated following a sudden quench across
a quantum critical point) may cut  the 
real axis for modes ${\vec k^*}$ such that  $|u_{\vec k^*}|^2=1/2$, at those instants of real time given by
\be
t_n^* = \frac 1 {2 \epsilon_{\vec{k}^*}^f}  \{ \pi(2n+1)\}.
\label{eq_time}
\ee
where ${\vec k}^*$ (i.e., $k_x^*$ and $k_y^*$) can take a continuous set of values in the thermodynamic limit.  At the instants of time when  the boundary points of the area corresponding to one sector of FZ (i.e., one particular $n$), determined by maximizing and minimizing   $\epsilon_{k^*}^f$, in Eq.~\eqref{eq_time} over
allowed values of ${\vec k}^*$,  touch the real time axis,  the time derivative of the rate function
of the return probability $I(t)$ 
\be
I(t)= - \int_{0}^{\pi} \frac{d^2 k}{(2\pi)^2} \ln \left(1 + 4 |u_{\vec k}|^2(|u_{\vec k}|^2-1) \sin^2 \epsilon_{\vec k}^f t \right).
\label{eq_rate_function}
\ee
exhibit cusp singularities. This scenario has been established analyzing the DQPTs those occur following a sudden quench of the staggered mass of the Haldane model
  from the non-topological phase to the topological phase \ct{vajna15} as long as the Haldane mass of the model is non-zero \ct{bhattacharya09}.

 Expressing  the Loschmidt overlap in Eq.~\eqref{eq_LO}  in the form   $L_{\vec k} (t) = |r_{\vec k} (t)| \exp(i {\phi_{\vec k} (t)})$, we find 
 \be
 \phi_{\vec k}(t) =\tan^{-1} \left\{(v_{\vec k}|^2 - |u_{\vec k}|^2) \tan (\epsilon_{\vec k}t)\right \}.
 \label{eq_phi1}
 \ee
To arrive at the gauge-independent form of the geometric phase  
 $\phi_{\vec k}^G(t)$, which is defined in the ray (density matrix)-space, we subtract from $\phi_{\vec k}$,   the corresponding dynamical phase  $\phi_{\vec k}^{\rm dyn} (t) = -\int_0^t ds \langle \psi_{0_{\vec k}}(s)|H_f|\psi_{0_{\vec k}}(s)\rangle = (|v_{ {\vec k}}|^2- |u_{\vec k}|^2) \epsilon_{\vec k}^f t$ in the form
 $\phi_{\vec k}^G(t) =\phi_{\vec k}(t) -\phi_{\vec k}^{\rm dyn}(t)$. With a purpose to define DTOP that characterises the DQPTs occurring  in two-dimensional models, we write: 
 
 \be
 \nu(t) =  \frac 1 {2\pi}  \oint_{EBZ} dk_x \frac {\partial}{\partial k_x} \left[ \oint dk_y\frac {\partial \phi_{\vec k}^G(t)}{\partial k_y} \right]
 \label{eq_dtop}
 \ee
 where $\nu(t)$ is the dynamical topological invariant or DTOP and the integration over $k_x$ spans over the effective Brillouin zone (EBZ) defined later. Although the exclusion of the dynamical phase from the total phase is essential to render $\phi_G$ gauge invariant, the closed integral of the partial derivative of the dynamical phase over the BZ along $k_y$ (and the EBZ along $k_x$) {vanishes} and does not contribute to the DTOP due to the absence of any singularity in the dynamical phase. We would also like to emphasize here that the DTOP (Eq.  \ref{eq_dtop}) defined here for two-dimensional systems is indeed a non-trivial generalisation of the DTOP proposed by \ct{budich15} for one dimensional systems.\\

{We emphasise that the PGP becomes singular at Fisher zeros as
the total phase $\phi_{\vec k}(t)$ in Eq. \ref{eq_phi1} is ill defined at critical times. Now, to define a DTOP, we consider a $k_y$ non-trivial homological circle at every point $k_x$ in the BZ and compute the closed integral $\oint dk_y\frac {\partial \phi_{\vec k}^G}{\partial k_y}$ over the same circle to obtain $\phi_{k_x}^G$. If $\phi_{\vec k}^G = \phi^G(k_x,k_y)$ is a discontinuous function at the boundary point of the $k_y$ circle, i.e, $\phi^G\left(k_x,k_y(l\rightarrow 0)\right)\neq\phi^G\left(k_x,k_y(l\rightarrow 1)\right)$ (where $l \in [0,1]$ is a parameter that parametrizes the homological circle along $k_y$ at each $k_x$), $\phi_{\vec k}^G$  displays some $2\pi$-discontinuous jumps along the $k_y$ circle yielding $\phi^G(k_x,k_y)= {2\pi\Theta(k_y-k_y^c)}+ {\tilde \phi}^G(k_x,k_y)$, where $k_y^c$ is the critical $k_y$ mode for a fixed $k_x$, $\Theta$ is the Heaviside step function, and ${\tilde \phi}^G(k_x,k_y)$ is a continuous function over the BZ. Thus, computing the derivative and performing the integration of $\frac {\partial \phi_{\vec k}^G}{\partial k_y}$ along the $k_y$ non-trivial homological circle of the torus counts the number of these $2\pi$ jumps as the $k_y$ circle gets spanned. $\phi_{k_x}^G = \oint dk_y\frac {\partial \phi_{\vec k}^G}{\partial k_y}$, as is easily observed, is pinned to zero at the TRIM points of the BZ (along $k_x$) and therefore, we again take the partial derivative of $\frac{1}{2\pi}\phi_{k_x}^G$ along $k_x$ and integrate this along the homological circle closed between two successive TRIM points along $k_x$ (thus defining an effective $k_x$ BZ), resulting in a topological number that changes at the boundaries of the real time axis where the FZs cross. Once again, this can be understood by expressing $\phi_{k_x}^G = \phi^G(k_x)= {2\pi\Theta(k_x-k_x^c)}+ {\tilde \phi}^G(k_x)$  where $k_x^c$ is the critical $k_x$ mode, and again ${\tilde \phi}^G(k_x)$ is a continuous function over the EBZ. The derivative detects the presence of a jump between $\phi^G(k_x)$ and $\phi^G(k_x+dk_x)$, and finally the integral over the effective $k_x$-BZ counts the total number of jumps giving rise to a quantised topological order parameter capable of detecting the DQPTs.}\

\begin{figure}
\includegraphics[height=2.5in, width=7.5cm]{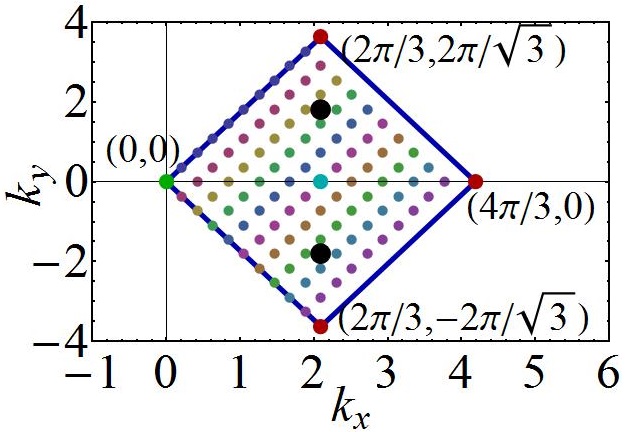}
\caption{(Color online) The rhomboidal Brillouin zone (BZ) that we
have chosen in this paper is illustrated for a $10 \times 10$ system. The Dirac points are shown in black (biggest dots) while other small dots represent the points
of the BZ. The four corner points and the highest symmetry point at the centre (with co-ordinates $(2\pi/3,0)$) marked by dots of medium size  are time reversal invariant momentum (TRIM) points where $M_H(\mathbf{k})=0$; while the left corner point $(0,0)$ (in green) and the central point (in cyan) are included in the BZ, other TRIM points (in red) are not. The EBZ along $k_x$ used in Eq. \eqref{eq_dtop} is actually along the $k_y=0$ line from $k_x=0$ to $k_x=\frac{2\pi}{3}$ whereas, the BZ along $k_y$ is the circle closed between $-\sqrt{3}k_x$ to $\sqrt{3}k_x$.}
\label{fig_BZ}
\end{figure}

\begin{figure*}[]
\centering
\subfigure[]{%
\includegraphics[width=.45\textwidth,height=6cm]{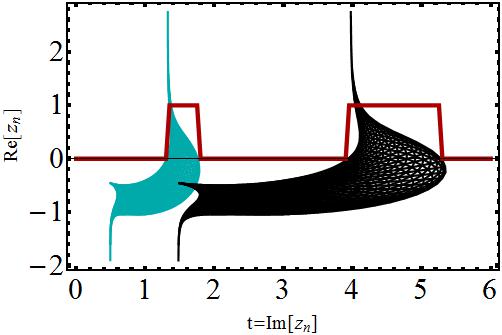}
\label{FZ_haldane}}
\hfill
\quad
\subfigure[]{%
\includegraphics[width=.45\textwidth,height=6cm]{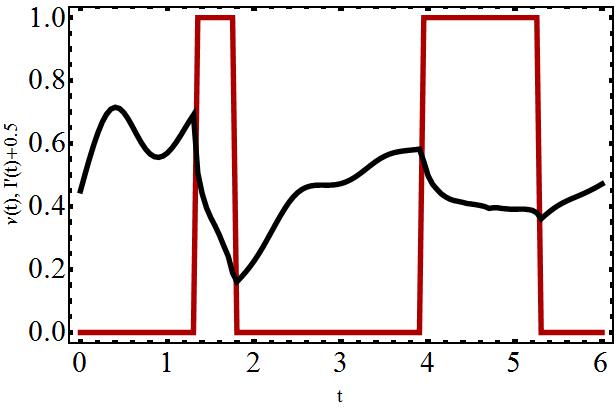}
\label{fig_non_anal}}
\caption{ (Color online) (a) Fisher zeros for sectors $n=0$ and $n=1$, in the complex $z$-plane following a sudden quench of the staggered mass of the Haldane model from the non-topological phase
to the topological phase. The FZs form areas which cut
the imaginary (real time) axis. The temporal evolution of the DTOP which is superimposed  on the same curve makes a discontinuous jump at the boundary point of the areas. (b) At these boundary points,
the derivative of the return probability $I'(t)$ (scaled by 0.5 to show the exact correspondence with $\nu(t)$) also exhibit non-analyticities (cusp singularities) clearly marking the presence of a DQPT.}
\end{figure*}

To illustrate the claim, we consider the 2D topological Haldane model \ct{haldane88} on a hexagonal lattice; within the periodic boundary condition one can work with the equivalent $2 \times 2$  Hamiltonian in the pseudospin basis parametried by quasi-momentum $\left(k_x,k_y\right)$ to yield, 

\begin{equation}
\begin{aligned}
\mathcal{H}(\vec{k}) & = 
\begin{pmatrix}
h_3(\vec{k}) & h_1(\vec{k})+ih_2(\vec{k}) \\
h_1(\vec{k})-ih_2(\vec{k}) & -h_3(\vec{k})
\end{pmatrix}
\end{aligned}
\label{eq_haldane_model}
\end{equation}

\noi where $h_3(\vec {k})= M + M_H({\vec k})$ with $M_H({\vec  k})$ being the quasi-momentum dependent Haldane mass and $M$ is the staggered (Semenoff) mass \ct{semenoff08}. We employ a sudden quenching
scheme in which the staggered mass is changed from an initial value $M_i=3$ to a final value $M_f=1$ so that the model, initially in the non-topological phase (with Chern number $C=0$), is quenched to the
topological phase with $C=0$ and analyze the subsequent temporal evolution of the DTOP defined in Eq. \eqref{eq_dtop} in connection to the DQPTs. (We refer to the Appendix B of ref.\onlinecite{bhattacharya09}
for an elaborate discussion on associated DQPTs.)

Referring to Fig. \ref{FZ_haldane}, we find the areas of FZs indeed cut the real axis; the boundary points for the sector $n=0$ touch the real time axis at $t_0^-$ and $t_0^+$ while, as predicted
by Eq. \eqref{eq_time}, boundary points corresponding to the sector $n=1$ occur at instant $t_1^-=3t_0^-$ and $t_2^+=3t_0^+$; the time-derivative $I'(t)$ also exhibit cusp singularities  at these instants $t_n^{\pm}$ as shown
in Fig. \ref{fig_non_anal}.  What is fascinating is that the DTOP jumps from zero to identity at every instant $t_n^{+}$ and jumps from identity to zero at every $t_n^-$. We reiterate that the instants $t_n^{-}(t_n^{+})$ are derived by maximizing (minimizing) $\epsilon_{k^*}$ over the allowed values of $k^*$ for a given $n$.

Question that needs to be addressed here concerns what causes the discontinuous jumps in the DTOP  of unit magnitude at $t_n^{-}$ and $t_n^{+}$. The DTOP essentially counts the singularities it encounters at each instant of time $t$. Thus, it stays at zero before the first singularity is encountered at $t_n^-$, where it jumps to $1$ and remains pinned at $1$ till it goes on encountering singularities upto $t_n^+$. Since, there are no singularities in the LOA within the time interval between $t_n^+$ and $t_{n+1}^-$ , the DTOP goes to zero and remains there till it again hits the next zone of singularities at $t_{n+1}^-$.

In short, we have proposed an emerging topological structure in the subsequent temporal evolution of a two dimensional quenched quantum system; this structure is characterised by a topological
invariant, namely, the  DTOP which is pinned to integer values and remarkably detects the singularities associated with corresponding DQPTs. Our claim is established studying the temporal
evolution and corresponding  DQPTs occurring in a two-dimensional topological Haldane model following a sudden quench from the non-topological phase to the topological phase where
$\nu(t)$ is pinned to either zero or unity. However,
it should be emphasized that the results we present are independent of the quenching scheme or the model.

\medskip

We acknowledge Shraddha Sharma for discussion. AD acknowledges SERB, DST for financial support.


\begin{thebibliography}{11}
 
 
 
 
 
\bi{bloch08} I. Bloch, J. Dalibard, and W. Zwerger, Rev. Mod. Phys. {\bf 80}, 885 (2008).
%
\bi{lewenstein12} M. Lewenstein, A. Sanpera, and V. Ahufinger, (Oxford University Press, Oxford (2012)).



\bibitem{jotzu14} G.  Jotzu,	M. Messer, R. Desbuquois,	M. Lebrat,	 T. Uehlinger,	D. Greif and  T.  Esslinger,    Nature
    {\bf 515},
    237  (2014).
    
\bi{greiner02} M. Greiner , O. Mandel, T.  W. Hansch and  I. Bloch, Nature {\bf 419}, 51 (2002).


\bi{kinoshita06} T. Kinoshita, T. Wenger and D. S. Weiss, Nature {\bf 440}, 900 (2006).



 \bi{gring12} M. Gring, M. Kuhnert, T. Langen, T. Kitagawa, B. Rauer, M. Schreitl, I. Mazets1, D. Adu Smith, E. Demler, and J. Schmiedmayer,
 Science {\bf 337}, 1318 (2012).
 
\bi{trotzky12}  S. Trotzky,	Y-A. Chen,	A. Flesch,	I. P. McCulloch,	U. Schollwck, J. Eisert and I. Bloch, Nature {\bf 8}, 325 (2012).

 \bi{cheneau12}  M. Cheneau,	P. Barmettler,	D.  Poletti,	 M. Endres,	P.  Schauss,	T. Fukuhara,	C. Gross,	I. Bloch,	C.  Kollath	 and S.  Kuhr, Nature {\bf 481}, 484  (2012).   



\bi{fausti11} D. Fausti, R. I. Tobey, , N. Dean,  S. Kaiser, A. Dienst, M. C. Hoffmann, S. Pyon, T. Takayama, H. Takagi,4, A. Cavalleri, Science {\bf 331}, 189 (2011). 

\bi{rechtsman13}  M. C. Rechtsman,	J. M. Zeuner,	Y.  Plotnik,	 Y.  Lumer,	D.Podolsky,	F.  Dreisow,	S. Nolte,	M. Segev	and  A. Szameit,    Nature
    {\bf 496}  196 (2013).
    
\bi{schreiber15}     M.  Schreiber,  S. S. Hodgman, P.  Bordia,  Henrik P. Lschen, M. H. Fischer, R. Vosk, E. Altman, U. Schneider, I. Bloch, Science {\bf 349}, 842 (2015).

\bi{calabrese06} P. Calabrese, and  J. Cardy,   Phys. Rev. Lett. {\bf 96}, 136801 (2006); J. Stat. Mech,
P06008 (2007).

\bi{rigol08} M. Rigol, V. Dunjko and M. Olshanii, \textit{Thermalization and its mechanism for generic isolated quantum systems}, Nature {\bf 452}, 854 (2008).



\bibitem{kitagawa10} T. Kitagawa, E. Berg, M. Rudner, and E. Demler,  Phys. Rev. B
{\bf 82}, 235114 (2010).

\bibitem{lindner11} N. H. Lindner, G. Refael and V. Galitski, Nat. Phys. {\bf 7}, 490-495, (2011). 


\bi{bermudez09} A. Bermudez, D. Patane,  L. Amico, M. A. Martin-Delgado,  Phys. Rev. Lett. {\bf 102}, 135702, (2009).

\bi{patel13} A. A. Patel, S. Sharma, A. Dutta,  Eur. Phys. Jour. B {\bf 86}, 367 (2013); A. Rajak and  A. Dutta,  Phys. Rev. E 89, 042125, 2014. P. D. Sacramento, \textit{Fate of Majorana fermions and Chern numbers after a quantum quench}, Phys. Rev. E {\bf 90} 032138, (2014); M. D. Caio, N. R. Cooper and M. J. Bhaseen,  Phys. Rev. Lett. {\bf 115}, 236403 (2015).    

\bibitem{thakurathi13} M. Thakurathi, A. A. Patel, D. Sen, and A. Dutta,  Phys. Rev. B 88, 155133 (2013).

\bibitem{mukherjee09}
V. Mukherjee V. and A. Dutta,  J. Stat. Mech. P05005 (2009).

\bibitem{das10}
A. Das,   Phys. Rev. B {\bf 82}, 172402 (2010).

\bibitem{Russomanno_PRL12}
{A. Russomanno,  A. Silva  and G. E. Santoro} , Phys. Rev.
  Lett. {\bf 109}, 257201 (2012); S. Sharma, A. Russomanno, G. E. Santoro and A. Dutta,  EPL {\bf 106},  67003 (2014).

\bi{bukov16} 
M. Bukov, L. D'Alessio and A. Polkovnikov,  Adv. Phys. {\bf 64} , No. 2, 139-226 (2016).

\bibitem{pal10} A Pal and D. A. Huse,  Phys. Rev. B {\bf 82}, 174411  (2010).

\bi{nandkishore15} R. Nandkishore, D. A. Huse,  Annual Review of Condensed Matter Physics,  {\bf 6}, 15-38 (2015).   

\bibitem {dziarmaga10} J. Dziarmaga,  Advances in Physics  {\bf 59}, 1063 (2010).

\bibitem{polkovnikov11} A. Polkovnikov, K. Sengupta, A. Silva, and M. Vengalattore, Rev. Mod. Phys. {\bf 83}, 863 (2011).	
		
	

\bi{dutta15} A. Dutta, G. Aeppli, B. K. Chakrabarti, U. Divakaran, T. 
Rosenbaum and D. Sen, \textit{Quantum Phase Transitions in Transverse Field 
Spin Models: From Statistical Physics to Quantum Information} (Cambridge 
University Press, Cambridge, 2015).

\bi{eisert15} J. Eisert, M. Friesdorf and C. Gogolin, Nat. Phys. {\bf 11}, 124 (2015).

\bi{alessio16} L. D'Alessio, Y.  Kafri, A. Polkovnikov, M. Rigol,  Adv. Phys. {\bf 65}, 239 (2016).
 
\bi{heyl13} M. Heyl, A. Polkovnikov, and S. Kehrein, Phys. Rev. Lett., {\bf 110}, 135704 (2013).

	
\bi {fisher65} M.E. Fisher, in {\it Boulder Lectures in Theoretical Physics} (University of Colorado, Boulder, 1965), Vol. 7.

\bi {lee52} C. Yang and T. Lee,  Phys. Rev. {\bf 87}, 404 (1952).

\bi{saarloos84} W. van Saarloos and D. Kurtze, J. Phys. A {\bf 17}, 1301 (1984).

\bibitem{peng16} X. Peng,  H. Zhou, B.-B. Wei,  J. Cui, J. Du, and R.-B. Liu,  Phys. Rev. Lett. {\bf 114}, 010601 (2015).	

%



\bi{karrasch13} C. Karrasch and D. Schuricht, Phys. Rev. B, {\bf 87}, 195104 (2013).

\bi{kriel14} N. Kriel, C. Karrasch, and S. Kehrein, Phys. Rev. B {\bf 90}, 125106 (2014).

\bi{heyl14} M. Heyl, Phys. Rev. Lett., {\bf 113}, 205701 (2014).


\bi{andraschko14} F. Andraschko, J. Sirker,  Phys. Rev. B {\bf 89}, 125120 (2014).

\bi{canovi14} E. Canovi, P. Werner, and M. Eckstein, Phys. Rev. Lett. {\bf 113}, 265702 (2014).

\bi{heyl15} M. Heyl,  Phys. Rev. Lett., {\bf 115}, 140602 (2015) .

\bi{palami15} T. Palmai, Phys. Rev. B {\bf 92}, 235433 (2015).

\bi{vajna14} S. Vajna and B. Dora, Phys. Rev. B {\bf 89}, 161105(R) (2014).

\bi{sharma15} S. Sharma, S. Suzuki and A. Dutta, Phys. Rev. B {\bf 92}, 104306 (2015).

\bi{vajna15} S. Vajna and B. Dora,  Phys. Rev. B {\bf 91}, 155127 (2015).

\bi{schmitt15} M. Schmitt and S. Kehrein,  Phys. Rev. B {\bf 92}, 075114 (2015).

\bi{budich15} J. C. Budich and  M. Heyl,  Phys. Rev. B 93, 085416 (2016).

\bi{sharma16} S. Sharma, U. Divakaran, A. Polkovnikov and A. Dutta,  Phys. Rev. B {\bf 93}, 144306 (2016).

\bi{divakaran16} U. Divakaran, S. Sharma and A. Dutta, Phys. Rev. E {\bf 93}, 052133 (2016).


\bi{huang16} Z.  Huang, and A.  V. Balatsky,  Phys. Rev. Lett. {\bf 117}, 086802 (2016).

\bi{puskarov16} T. Puskarov and D. Schuricht,  arXiv: 1608.05584 (2016).

\bi{zhang16} J. M. Zhang abd  H.-T. Yang,  arXiv: 1605.05403 (2016).

\bi{heyl16} M. Heyl, arXiv: 1608.06659 (2016).



\bi{pollmann10} F. Pollmann, S. Mukerjee, A. G. Green, and J. E. Moore,  Phys. Rev. E {\bf 81}, 020101(R) (2010).









\bi{quan06} H.T. Quan, Z. Song, X.F. Liu, P. Zanardi, and C.P. Sun,  Phys.Rev.Lett. {\bf 96}, 140604 (2006).

\bi{rossini07} D. Rossini, T. Calarco, V. Giovannetti, S. Montangero, R. Fazio, Phys. Rev. A {\bf 75}, 032333 (2007).

\bi{cucchietti07} F. M. Cucchietti, $et~al$,  Phys. Rev. A {\bf 75}, 032337 (2007);
 C. Cormick and J. P. Paz,  Phys. Rev. A {\bf 77}, 022317 (2008).


\bi{venuti10} Lorenzo C Venuti and P. Zanardi, Phys. Rev. A {\bf 81}, 022113 (2010); 
Lorenzo C. Venuti, N. T. Jacobson, S. Santra, and P.  Zanardi,  Phys. Rev. Lett. {\bf 107}, 010403 (2011).

\bi{sharma12} S. Sharma, V. Mukherjee, and A. Dutta,  Eur. Phys. J. B, {\bf 85}, 143 (2012).



\bi{mukherjee12} V. Mukherjee, S. Sharma, A. Dutta, Phys. Rev. B {\bf 86}, 020301 (R) (2012).

\bi{nag12} T. Nag, U. Divakaran and A. Dutta,  Phys. Rev. A {\bf 93}, 012112(2016)




\bi{dora13} B. Dora, F. Pollmann, J. Fortgh, G. Zarand, Phys. Rev. Lett. {\bf 111}, 046402 (2013);  R. Sachdeva, T. Nag, A. Agarwal, A. Dutta,  Phys. Rev. B {\bf 90}, 045421 (2014).





\bi{gambassi11} A. Gambassi and  A.  Silva,  arXiv: 1106.2671 (2011);  Phys. Rev. Lett. {\bf 109}, 250602 (2012); P. Smacchia and A. Silva, Phys. Rev. E {\bf 88}, 042109, (2013); A. Russomanno, S. Sharma, A. Dutta and G. E. Santoro, J. Stat. Mech., P08030  (2015).

\bi{zanardi07_echo} P. Zanardi. H. T. Quan, X. Wang and C. P. Sun, Phys. Rev. A {\bf 75}, 032109 (2007).

\bi{dorner12} R. Dorner, J. Goold, C. Cormick, M. Paternostro and V. Vedral,  Phys. Rev. Lett. {\bf 109}, 160601 (2012);S. Sharma and A. Dutta, Phys. Rev. E {\bf 92}, 022108 (2015).

\bi{bhattacharya09} U. Bhattacharya and A, Dutta, arXiv: 1610.02674 (2016).

\bi{flaschner16} N. Flaschner, D. Vogel, M. Tarnowski, B, S. Rem, D.-S. Luhmann, M. Heyl, J.  Budich, L. Mathey, K. Sengstock, C. Weitenberg, 
arXiv:1608.05616 (2016).

\bibitem{haldane88}
M. Haldane,  
\newblock Phys. Rev. Lett. {\bf 61}, 18, (1988).


    
\bi{semenoff08} G. W. Semenoff, V. Semenoff, and Fei Zhou, 
Phys. Rev. Lett. {\bf 101}, 087204







\end{thebibliography}
\end{document}